\documentclass[10pt]{article}

\usepackage{nebius}

\usepackage[utf8]{inputenc} %
\usepackage{url}            %
\usepackage{graphicx}       %
\usepackage{booktabs}       %
\usepackage{amsmath}        %
\usepackage{amsfonts}       %
\usepackage{nicefrac}       %
\usepackage{colortbl}       %
\usepackage{tikz}           %
\usetikzlibrary{arrows.meta,positioning,calc}
\usepackage{pgfplots}       %
\pgfplotsset{compat=1.18}
\usepackage{fontawesome5}   %
\usepackage{hyperref}       %
\nebiushypersetup
\hypersetup{
  pdftitle={SpectralBridge: Adding High-Frequency Capability to Large Audio-Language Models},
  pdfauthor={Sergei Polezhaev, Saba Anjum, Michael Gormish},
  pdfsubject={Preprint},
}

\newcommand{\methodname}{\textsc{SpectralBridge}}

\newcommand{\Proj}{\mathcal{P}}
\newcommand{\Lowpass}{\operatorname{LP}}

\providecommand{\nebiussans}{\sffamily}
\providecommand{\nebiussemibold}{\sffamily\fontseries{semibold}\selectfont}

\colorlet{lowbandcolor}{nebiusdeepblue}
\colorlet{lowbandfill}{nebiuslightblueii}
\colorlet{highbandcolor}{nebiusviolet}
\colorlet{highbandfill}{nebiuslavenderii}
\colorlet{nullcolor}{nebiusdeepblue!45!white}
\colorlet{nullfill}{nebiusdeepblue!6!white}
\colorlet{frozencolor}{nebiusdeepblue!35!white}
\colorlet{frozenfill}{nebiuslightblue}
\colorlet{frozenicon}{nebiusdeepblue!60!white}
\colorlet{trainedcolor}{nebiusdeepblue}
\colorlet{trainedfill}{nebiuslime}
\colorlet{correctfill}{nebiuslime}
\colorlet{wrongfill}{nebiuslavenderii}
\colorlet{oursshade}{nebiuslightblueii}
\colorlet{nebiusgray}{nebiusdeepblue!45!white}
\colorlet{nebiuslightgray}{nebiusdeepblue!12!white}
\colorlet{nebiushairline}{nebiusdeepblue!30!white}

\newcommand{\nebiusfont}{\nebiussans}

\pgfplotsset{
  nebius axis/.style={
    axis background/.style={fill=white},
    axis x line*=bottom,
    axis y line*=left,
    axis line style={draw=nebiusdeepblue!35,line width=0.4pt},
    y axis line style={draw=none},
    tick style={draw=nebiusdeepblue!35,line width=0.4pt},
    ytick style={draw=none},
    tick label style={font=\nebiusfont\scriptsize,text=nebiusdeepblue},
    x tick label style={align=center},
    label style={font=\nebiusfont\scriptsize,text=nebiusdeepblue},
    title style={font=\nebiussemibold\scriptsize,text=nebiusdeepblue},
    major tick length=2pt,
    ymajorgrids=true,
    grid style={draw=nebiusdeepblue!12,line width=0.4pt},
    legend style={font=\nebiusfont\scriptsize,draw=none,fill=none},
    every node near coord/.style={font=\nebiusfont\scriptsize,text=nebiusdeepblue,fill=none,inner sep=0pt},
    /pgf/number format/assume math mode=true,
  },
  nebius image axis/.style={
    nebius axis,
    axis x line*=box,
    axis y line*=box,
    axis line style={draw=nebiusdeepblue!35,line width=0.4pt},
    y axis line style={draw=nebiusdeepblue!35,line width=0.4pt},
    ytick style={draw=nebiusdeepblue!35,line width=0.4pt},
    tick align=outside,
    xtick pos=bottom,
    ytick pos=left,
    ymajorgrids=false,
    enlargelimits=false,
    axis on top,
  },
}

\tcbset{
  nebiusprompt/.style={
    breakable,
    enhanced,
    colback=nebiuslightblue,
    colframe=nebiuslightblue,
    boxrule=0pt,
    arc=2pt,
    left=7pt, right=7pt, top=5pt, bottom=5pt,
    borderline west={2pt}{0pt}{nebiusdeepblue},
  },
}
\newtcolorbox{nebiusexample}{nebiusprompt, before skip=6pt, after skip=6pt}

\begin{document}

\thispagestyle{empty}
\nebiuslogo

\begin{nebiustitlebox}
  \centering

  \vspace{0.3em}
  {\nebiustitlefont SpectralBridge: Adding High-Frequency Capability\\ to Large Audio-Language Models\par}
  \vspace{1.1em}

  {\nebiusauthorfont
    \textbf{Sergei Polezhaev}\textsuperscript{1*},\
    \textbf{Saba Anjum}\textsuperscript{1},\
    \textbf{Michael Gormish}\textsuperscript{1}\par
  }
  \vspace{0.5em}

  {\nebiusaffilfont
    \textsuperscript{1}Nebius AI\\[0.3em]
    \textsuperscript{*}Correspondence to: spolezhaev@nebius.com\par
  }
  \vspace{0.6em}

\begin{abstract}
Large audio-language models (LALMs) have become highly capable on a wide range of tasks spanning speech,
 music, and environmental sound. However, many modern LALMs resample input audio to 16~kHz mono,
making all frequencies above 8~kHz unobservable. We present \methodname, a
lightweight extension that gives such a model access to the 8--16~kHz band by joining low- and
upper-band representations with time-aligned paired cross-frequency attention.
Applied to Audio Flamingo Next, the method trains only a small adapter and rank-8 LoRA modules, while leaving
the original audio encoder, audio projector, and base language model frozen.
\methodname{} improves both zero-shot transfer and downstream fine-tuning: it raises zero-shot
balanced accuracy on the ESC-50 audio classification dataset from 75.5 to 90.5, and when fine-tuned on Insect-10, a 10-way
insect species task we construct from InsectSet459 recordings,
under the same budget as a LoRA fine-tuned baseline, it reaches 55.4 balanced
accuracy against the baseline's 26.2. On a benchmark where the answer depends only on the
8--16~kHz band, it gains 30 mAP over the frozen model. Erasing the upper-band
stream from the trained model removes these gains, showing that they come from
the high-frequency signal itself rather than from extra tokens or the adapted
prompt format.
\end{abstract}
\end{nebiustitlebox}

\section{Introduction}
Large audio-language models (LALMs) place increasingly capable language models
behind audio encoders, allowing a common interface for understanding speech,
music, environmental sounds, and other audio~\citep{gong2023ltu,tang2023salmonn,chu2023qwenaudio,kong2024audioflamingo,ghosh2026afnext}.
Recent systems have expanded the scale of training, the range of tasks, and the
duration of audio that can be processed. However, a basic constraint has often been overlooked: which parts of the acoustic signal reach the model input at all.
Traditionally, many LALMs inherit Automatic Speech Recognition (ASR)-oriented front ends. Converting input
audio to 16~kHz mono is a reasonable engineering choice for speech-heavy
applications, since much of the linguistic information in speech lies below
8~kHz, and removing higher frequencies reduces compute and even suppresses some
irrelevant noise.

For more general audio understanding, this restriction, however, is not harmless. Many
environmental sounds, musical instruments, and recording artifacts carry
important evidence in the upper frequency range. Bells, cymbals, insects,
sibilance, hiss, and the ``air'' or brightness of a music mix often depend on
structure above 8~kHz. For example, a music co-producer or mastering assistant that cannot
hear this range may still be helpful, but it cannot reliably answer
questions about top-end detail.

The Audio Flamingo Next (AF-Next) model~\citep{ghosh2026afnext}, for instance, uses an AF-Whisper encoder and first resamples its input
to 16~kHz mono, as does the underlying Whisper model family~\citep{radford2022whisper}.
By the Nyquist limit~\citep{shannon1949communication}, the resulting samples cannot represent frequencies above 8~kHz. To avoid aliasing of the high-frequency content into the low-frequency band, the resampler applies a low-pass filter to the signal. After low-pass filtering and resampling, no amount of language-model reasoning can recover information from the high-frequency signal.

It is important to note that earlier Audio Flamingo models used higher-bandwidth
audio encoders~\citep{kong2024audioflamingo,ghosh2025audioflamingo2}. Starting
with Audio Flamingo 3~\citep{goel2025audioflamingo3}, however, the authors switched to a unified AF-Whisper
encoder to improve training stability. This encoder is initialized from a Whisper
Large-v3 model~\citep{radford2022whisper} and then further trained on large,
diverse speech, sound, and music data. Since this produces a strong and stable
audio representation, it is preferable to reuse the frozen encoder rather than
replace or retrain it from scratch. This creates the central question of this
work: can we keep the frozen AF-Whisper encoder, but still expose it to
high-frequency evidence?

The lack of high frequencies is easy to overlook on conventional semantic benchmarks. A bell, cymbal, or insect recording can sometimes be recognized from correlated
evidence below 8~kHz, even when its upper partials or noise components are
informative. Good accuracy on the original recording therefore does not
establish that a model can use the upper band.

We address this problem with \methodname, a parameter-efficient extension of
AF-Next. Given wideband audio, we retain the ordinary 0--8~kHz view and
translate the 8--16~kHz band into 0--8~kHz with time-preserving heterodyning.
Both views are passed through the same frozen AF-Whisper encoder. At each time index,
a small paired cross-attention adapter interacts only across the two frequency
views; it never attends to another time position. A high-minus-null
representation and an energy-dependent gate make the residual exactly inactive
when upper-band input is removed or contains very little energy. The two projected streams are then supplied
to the AF-Next language model, whose attention layers are adapted with LoRA.
While this architecture produces an additional $N$ audio tokens, we also study
ways to reduce the number of tokens back to $N$ without losing quality.

On the held-out semantic test, opened once after freezing the architecture,
checkpoint rule, seeds, prompts, metrics, and bootstrap procedure,
\methodname{} obtains 81.4 mAP and 69.7\% balanced accuracy. It improves over a matched simple two-view LoRA baseline by 13.4 mAP
and 9.0 balanced-accuracy points, and over frozen AF-Next by 30.1 mAP. The
same trained weights fall to 49.4 mAP when the upper-band view is replaced by
digital silence. Turning off only the learned paired residual also degrades
development-set performance, isolating a benefit from local
cross-frequency interaction beyond access to a second stream.

The contributions of this paper are as follows:
\begin{itemize}
    \item We develop \methodname, a lightweight method that adds a second upper-band view to an AF-Next model by providing a time-aligned heterodyned representation of both low-frequency and high-frequency views to a frozen AF-Whisper encoder and a paired cross-frequency attention adapter.
    \item We collect a training dataset that teaches the model to use upper-band information. The dataset is constructed so that most examples are synthetic or generic signal-physics questions, and it does not overlap in task or sound class with the downstream zero-shot evaluations.
    \item We show that \methodname{} significantly improves zero-shot semantic performance on a held-out test set, and that the model's answers are causally affected by the upper-band evidence.
    \item We also show that fine-tuning \methodname{} on downstream high-frequency tasks improves performance over fine-tuning the baseline AF-Next model under the same budget.
\end{itemize}

\section{Related Work}
\label{sec:related_work}

\subsection{Large Audio-Language Models}

Audio-language models connect an acoustic representation to a language model,
enabling instruction following and open-ended question answering over sound.
Recent examples include LTU, SALMONN, Qwen-Audio, Qwen2-Audio, and the
Audio Flamingo family~\citep{gong2023ltu,tang2023salmonn,chu2023qwenaudio,chu2024qwen2audio,kong2024audioflamingo}.
For example, Audio Flamingo Next was designed to perform well on multiple audio domains such as speech, music, and general sound recognition. It maps mono
16~kHz log-mel features through the AF-Whisper encoder, projects the resulting
sequence into the Qwen 2.5-7B language model, and uses timestamp-aware rotary
embeddings for long audio~\citep{ghosh2026afnext}. We use AF-Next as the base
model because, at the time of this work, it is one of the strongest open-source
models covering multiple audio domains and tasks, including music and general
sound, while its 16~kHz input path makes the high-frequency limitation especially clear.
\subsection{Adapting Pretrained Audio Models to Wider Bandwidths}

The bandwidth mismatch between pretrained audio models and higher-sampling-rate
downstream audio has recently received increasing attention. Closest to our
setting, \citet{fang2026amsa} propose Anchored Multi-Sampling-Rate Adaptation
(AMSA). AMSA partitions a wideband Short-Time Fourier Transform (STFT) representation into
frequency subbands and maps each subband to an in-band log-mel representation
compatible with a spectrogram Transformer pretrained at 16~kHz. A canonical
16-kHz branch is retained as an anchor, while auxiliary subband branches expose
the model to information from additional frequency ranges. The authors further
show that naive fusion of heterogeneous subbands can be difficult to optimize
and introduce separate branch supervision before final aggregation.

Other work instead changes the acoustic frontend itself. MSR-HuBERT replaces
HuBERT's fixed-rate convolutional frontend with sampling-rate-specific
downsampling branches, allowing signals between 16 and 48~kHz to be mapped to a
common temporal resolution without first resampling away their high-frequency
content~\citep{huang2026msrhubert}. This approach targets multi-rate
self-supervised representation learning and requires modifying and pretraining
the acoustic frontend. In contrast, we ask whether the unchanged, frozen
16-kHz encoder of an existing LALM can be reused to represent a previously
inaccessible frequency band.

Frequency translation provides another way to reuse a pretrained model outside
its original spectral range. \citet{svendsen2026seal} apply a pretrained Audio
Spectrogram Transformer (AST) to ultrasonic acoustic-emission measurements for
hydraulic wear classification. Their preprocessing
isolates a 50--100~kHz band and heterodynes it toward lower frequencies before
low-pass filtering and resampling to the 16-kHz input expected by AST. This
provides direct precedent for translating out-of-band physical measurements
into the passband of a pretrained audio network. Their objective, however, is a
task-specific three-class industrial classifier, and the translated signal
replaces the ordinary audio input. Our setting instead preserves the original
low-band representation, introduces the translated upper band as
\emph{additional} evidence, and studies whether a generative LALM can learn to
reason over that evidence for semantic concepts not used during training.

Some LALMs have also combined encoders operating at different bandwidths.
TinyGiantALM, for example, uses a 16-kHz Whisper stream together with a
48-kHz HTS-AT stream and an additional CLAP representation before language-model
reasoning~\citep{ly2026tinygiant}. This gives the language model access to a
high-bandwidth encoder, but also changes the representation family. Our preliminary
experiments suggest that when the added high-band encoder is weaker than the
frozen encoder, even if it was trained natively on full-band audio, the final
system can be limited by that representation.

\section{Method}
\label{sec:method}

\begin{figure}[t]
    \centering
    \resizebox{0.92\linewidth}{!}{\input{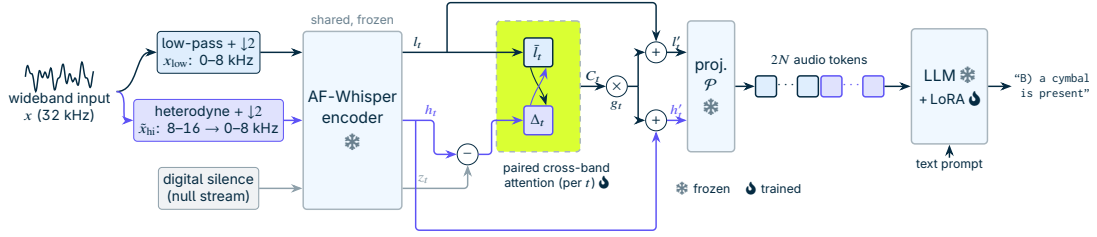}}
    \vspace{-1.5mm}
    \caption{Overview of \methodname. The 32~kHz input is split into the stock
    low-band view and an 8--16~kHz view heterodyned into the encoder passband;
    both, plus a silence null stream, pass through the shared frozen AF-Whisper
    encoder. The trained adapter attends across the two per-timestamp frequency
    tokens and adds the gated result back to both streams; the frozen projector
    yields the $2N$-token audio prefix for the frozen LLM with trained LoRA.}
    \label{fig:architecture}
    \vspace{2mm}
\end{figure}

\subsection{AF-Next Processing}
Figure~\ref{fig:architecture} gives an overview of the full \methodname{}
pipeline. Let $x \in \mathbb{R}^{S}$ be a mono waveform sampled at
$f_s = 32$~kHz. The stock AF-Next preprocessing first resamples the audio to
16~kHz. To do this, it applies an anti-aliasing low-pass filter and then keeps
every second sample:
\begin{equation}
    x_{\mathrm{low}}^{(\mathrm{full})}[n] = (x * h_{\mathrm{lp}})[n],
    \qquad
    x_{\mathrm{low}}[m] = x_{\mathrm{low}}^{(\mathrm{full})}[2m],
    \qquad m = 0, \dots, \lfloor S/2 \rfloor - 1.
    \label{eq:stock_input}
\end{equation}
Here $h_{\mathrm{lp}}$ is the low-pass filter and $*$ denotes convolution. The
filter keeps frequencies below about 8~kHz and suppresses higher frequencies
before we drop every other sample. This filtering is necessary because otherwise
the removed high-frequency content would fold, or alias, into the 0--8~kHz band
after downsampling. As a consequence, the waveform sent to the stock encoder
preserves the 0--8~kHz band, while the original 8--16~kHz content is discarded
before the audio reaches the encoder.

After that, the AF-Whisper feature extractor computes
log-mel features $M_{\mathrm{low}} = \operatorname{LogMel}(x_{\mathrm{low}})$, the frozen audio encoder maps them to
hidden states $H_{\mathrm{low}} = f_a(M_{\mathrm{low}})$, and then the projector converts them into language-model embedding space:
\begin{equation}
    A_{\mathrm{stock}} = \left[ \Proj(H_{\mathrm{low},1}), \ldots, \Proj(H_{\mathrm{low},N}) \right].
    \label{eq:afnext}
\end{equation}
Since $x_{\mathrm{low}}$ is low-pass filtered to 8~kHz, any two waveforms that differ only above 8~kHz give the same $A_{\mathrm{stock}}$.

\subsection{High-Band View by Frequency Translation}

To enable processing of the high-band part of the signal, we first extract it from the original 32~kHz waveform by subtracting the full-rate low-pass filtered version:
\begin{equation}
    x_{\mathrm{hi}} = x - x_{\mathrm{low}}^{(\mathrm{full})}.
\end{equation}
We reuse the original encoder that was trained to accept only 0--8~kHz content, so we translate the 8--16~kHz band down to 0--8~kHz by heterodyning:
\begin{equation}
    \tilde{x}_{\mathrm{hi}}[n]
    =
    D_2\!\left(
        \Lowpass_{8}
        \left(
            2\,x_{\mathrm{hi}}[n]\cos\left(2\pi \frac{8000}{32000}n\right)
        \right)
    \right).
    \label{eq:heterodyne}
\end{equation}
Here $D_2$ denotes decimation by a factor of two. Multiplication by the cosine shifts a sinusoidal component at frequency $f$ to
components at $|f-8|$ and $f+8$~kHz. The low-pass filter removes the upper image,
so content from 8--16~kHz is translated to 0--8~kHz while
keeping its temporal alignment with the low-band stream.

Both views are then passed independently through the AF-Whisper
feature extractor and encoder:
\begin{equation}
    M_{\mathrm{hi}} = \operatorname{LogMel}(\tilde{x}_{\mathrm{hi}}), \qquad
    H_{\mathrm{hi}} = f_a(M_{\mathrm{hi}}).
\end{equation}
We also compute a null stream
\begin{equation}
    M_{0} = \operatorname{LogMel}(0), \qquad
    Z = f_a(M_{0}),
\end{equation}
using digital silence with the same duration and mask. The null stream is used
to represent what the frozen encoder emits in the absence of upper-band input.
This makes the high-frequency evidence at time $t$ a literal difference from
the empty input rather than an unconstrained second embedding.

\subsection{Paired Cross-Band Adapter}

Let $l_t,h_t,z_t \in \mathbb{R}^{d}$ be the low, high, and null AF-Whisper
hidden states at time index $t$. We first normalize the low stream and the
high-minus-null term,
\begin{equation}
    \bar{l}_t = \mathrm{RMSNorm}_{L}(l_t), \qquad
    \Delta_t =
    \mathrm{RMSNorm}_{H}(h_t)-\mathrm{RMSNorm}_{H}(z_t).
    \label{eq:delta}
\end{equation}
The adapter treats each timestamp as a two-token sequence
$R_t=[\bar{l}_t;\Delta_t] \in \mathbb{R}^{2 \times d}$ and applies multi-head
self-attention only along this two-element frequency axis:
\begin{equation}
    C_t =
    \operatorname{MHA}(R_t)
    =
    \operatorname{concat}_{m=1}^{M}
    \left[
    \operatorname{softmax}
    \left(
        \frac{R_tW_Q^{(m)}(R_tW_K^{(m)})^\top}{\sqrt{d_a/M}}
    \right)
    R_tW_V^{(m)}
    \right]W_O .
    \label{eq:paired_attention}
\end{equation}
This module accepts only these two views at the same time index, so it cannot attend to other time positions. 

The paired residual is gated by the amount of high-minus-null delta:
\begin{equation}
    g_t = \frac{\|\Delta_t\|_2^2/d}{1+\|\Delta_t\|_2^2/d},
    \qquad
    [l'_t;h'_t] = [l_t;h_t] + g_t C_t .
    \label{eq:gate}
\end{equation}
The output projection $W_O$ is initialized with zeros, so at the start of training the hidden states remain unchanged. Moreover, if the high view is replaced by digital silence, then
$h_t=z_t$, $\Delta_t=0$, and $g_t=0$. Therefore the paired residual is exactly
inactive for the high-frequency-erased control, even after training.

The projected audio streams are then concatenated:
\begin{equation}
    A_{\methodname}
    =
    \left[
        \Proj(l'_1),\ldots,\Proj(l'_N);
        \Proj(h'_1),\ldots,\Proj(h'_N)
    \right].
    \label{eq:spectralbridge_tokens}
\end{equation}
The AF-Whisper encoder, the AF-Next projector, and the base language model are
kept frozen. We train the paired adapter and rank-8 LoRA modules~\citep{hu2022lora}
on the attention projections of all language-model layers. The model is optimized with a standard cross-entropy objective over the answer tokens.

\section{Training Dataset}
\label{sec:dataset}

To our knowledge, there are no open datasets focused specifically on high-frequency audio understanding. Although datasets exist in which high-band content is necessary or important for correct answers, isolating such examples from larger collections is difficult. Moreover, for zero-shot evaluation we require the training data to exclude any task formulation similar to the test prompts, to avoid potential data leakage.

We therefore construct a synthetic training dataset. It consists of
multiple-choice signal-physics questions over synthetic textures and filtered
recordings from AudioSet-derived clips~\citep{gemmeke2017audioset}. All
audio used for training and evaluation is resampled to 32~kHz before processing.
Labels are generated by deterministic transformations, including bandwidth changes,
relative upper-band gain, cross-band synchronization, and phase or coherence modifications. We balance answer letters and
option orderings within each prompt family, hold out disjoint development
paraphrases, and include low-band-only questions with high-band distractors. Examples and more details about the training dataset are presented in Appendix~\ref{app:dataset_examples}.

\subsection{Downstream Fine-Tuning Data}
We also evaluate the performance of \methodname{} on downstream fine-tuning
tasks that may benefit from high-band information.

\paragraph{Insect-10.} We build a 10-way insect species classification task
from the InsectSet459~\citep{faiss2025insectset459} validation audio. We select
ten well-covered species (five cicadas and five crickets or katydids), use
10-second excerpts rather than full recordings, and split them into
observation-disjoint train, development, and test sets. 

\paragraph{CEB bird pairs.} We additionally evaluate a CEB train-XenoCanto~\citep{ceb2026xenocanto} bird-pair task: 18 screened binary species-pair tasks with 2,958
training, 816 development, and 508 test rows.

Models fine-tuned on any
downstream suite are excluded from the corresponding zero-shot claims.

\section{Evaluation}
\label{sec:evaluation}

\subsection{Baselines and Variants}

To study the effect of each component, we consider several baselines.
The frozen baseline is the public AF-Next model that receives only
$A_{\mathrm{stock}}$. The low-band LoRA baseline receives the same low-band
tokens but trains the language-model LoRA weights; any gain from this
model cannot be due to high-frequency evidence. The simple two-view LoRA baseline
concatenates $\Proj(H_{\mathrm{low}})$ and $\Proj(H_{\mathrm{hi}})$ without paired cross-band attention, and trains the same LoRA modules.

The final method uses $2N$ audio tokens. To study whether this extra token budget
is necessary, we also consider exact-$N$ variants. In the half-rate variant, the
paired adapter is applied first and then each band is pooled within adjacent
two-frame windows. For a stream $U=(u_1,\ldots,u_N)$ and window
$B_i=\{2i-1,2i\}$, the pooled token is
\begin{equation}
    p_i(U) = \sum_{j \in B_i} \alpha_{ij} u_j, \qquad
    \alpha_{ij}
    =
    \frac{\exp(s(u_j))}{\sum_{k \in B_i}\exp(s(u_k))}.
\end{equation}
The scoring functions are initialized at zero, making the initial behavior
exact adjacent averaging. The low stream keeps $\lceil N/2\rceil$ pooled tokens
and the high stream keeps $\lfloor N/2\rfloor$, so the language model receives
exactly $N$ audio tokens while the two bands remain separate.

All evaluations use a multiple-choice format. For each prompt we score every answer
option by the log-likelihood of its answer letter, and average scores over
three random option orderings to reduce position bias. For binary and
multiclass classification we report balanced accuracy (BAcc), i.e. the mean
per-class accuracy. For binary target-presence tasks we also report mean
average precision (mAP), which measures threshold-free ranking quality. For
paired counterfactual datasets we report paired rank: the fraction of matched
pairs where the truly positive member receives a higher positive score than
the matched negative member.

\subsection{Counterfactual High-Band Evaluation}

We evaluate the model on a zero-shot benchmark that is constructed from FSD50K~\citep{fonseca2022fsd50k}. However, we do not use the original
sound-classification task. Instead, we construct pairs where both examples share the same low-band carrier
recording. The two members differ only in the 8--16~kHz additional signal: one
receives a donor containing the target class, such as a bell, cymbal or insect,
and the other receives an energy-matched non-target donor. Low-band content,
overall duration, carrier identity, and loudness are matched inside the pair.
Therefore, a model can solve the task only by using upper-band semantic evidence,
not by recognizing the carrier recording.

This design is complemented by a same-weight high-frequency-erased control evaluation. We
evaluate a trained model again after replacing the high stream with digital
silence, while preserving the checkpoint, prompt, token count, timestamps, and
masks. A performance
drop in this control is therefore evidence that the answer depends causally on
the 8--16~kHz waveform.

\subsection{Natural Transfer}

We additionally evaluate on natural, externally meaningful datasets. InsectSet459~\citep{faiss2025insectset459} tests broad insect-order recognition from field
recordings, where many clips contain substantial energy above 8~kHz.
LibriSeVoc~\citep{sun2023librisevoc} tests whether genuine speech is ranked
above neural-vocoder resyntheses, whose upper band is often imperfect.
ESC-50~\citep{piczak2015esc} and TinySOL~\citep{cella2020tinysol} test multiclass
environmental-sound and instrument recognition. None of these corpora or task
types appear in the generic training dataset. Representative prompts for
these evaluation datasets are listed in Appendix~\ref{app:dataset_examples}.

\subsection{Protocol}
Training uses three
seeds, 600 optimization steps, effective batch size 8, a OneCycle learning-rate
schedule, and learning rate $10^{-4}$. Downstream fine-tuning uses 300 steps
with the same effective batch size. We initialize from the authors' Hugging Face
release of Audio Flamingo Next, \texttt{nvidia/audio-flamingo-next-hf}
~\citep{nvidia2026audioflamingonexthf}, whose license allows non-commercial
research use. Experiments were run on an internal GPU cluster node with 8
NVIDIA H100 80GB HBM3 GPUs. The main training and fine-tuning jobs used four
H100s per run, with different seeds and ablations parallelized across the node.
The audio encoder, stock audio projector, and base language model remain frozen;
trained parameters are the paired adapter, the optional exact-$N$ pooling module,
and LoRA weights. The exact-$N$ half-rate student is initialized from the
trained $2N$ paired model, whose adapter and LoRA weights are then kept frozen;
only the pooling module (about 5k parameters) is trained for 50 steps on the
same dataset with the task cross-entropy loss plus a distillation term on
the teacher's answer-letter logits.

\section{Results}
\label{sec:results}

\subsection{Zero-Shot Counterfactual Transfer}

Table~\ref{tab:constructed_fsd_test} shows the sealed constructed-FSD test
result, evaluated once after the protocol was frozen. Frozen AF-Next performs around chance, as
expected, because the target evidence is absent from its input. The simple
two-view LoRA baseline improves over the frozen model, showing that exposing
a translated high stream helps. The paired cross-band adapter gives the largest
gain, reaching 81.4 mAP and 80.0 paired-rank accuracy. When the same trained
weights are evaluated with the high stream erased, mAP falls back to 49.4,
confirming that the improvement is carried by upper-band evidence.

\begin{table}[t]
    \centering
    \caption{Constructed-FSD sealed test. Values are percentages with $\pm$ reporting SEM across seeds. The dataset was constructed to ensure that the task cannot be solved without using upper-band evidence, which is why the Low-band LoRA baseline performs at chance, as does the Frozen AF-Next.}
    \label{tab:constructed_fsd_test}
    \small
    \begin{tabular}{lccc}
        \toprule
        Model & BAcc & mAP & Paired rank \\
        \midrule
        Frozen AF-Next & $51.3 \pm 1.2$ & $51.3 \pm 1.2$ & $45.0 \pm 2.1$ \\
        Low-band LoRA & $51.2 \pm 1.1$ & $51.4 \pm 0.9$ & $45.2 \pm 1.8$ \\
        Simple two-view LoRA & $60.7 \pm 1.1$ & $68.0 \pm 1.3$ & $72.5 \pm 1.4$ \\
\rowcolor{oursshade}
\textbf{\methodname{} paired 2N} & $\mathbf{69.7 \pm 1.0}$ & $\mathbf{81.4 \pm 0.8}$ & $\mathbf{80.0 \pm 1.1}$ \\
        \methodname{}, high erased & $49.7 \pm 1.3$ & $49.4 \pm 1.4$ & $45.6 \pm 2.2$ \\
        \bottomrule
    \end{tabular}
\end{table}

\begin{table}[t]
\centering
\caption{Real multiclass zero-shot benchmarks. Values are balanced accuracy; $\pm$ reports SEM across seeds.}
\label{tab:multiclass}
\small
\begin{tabular}{lcc}
\toprule
 Method & ESC-50 10-way & TinySOL 14-way \\
\midrule
Frozen AF-Next   & $75.5 \pm 0.8$ & $13.4 \pm 0.5$ \\
Low-band LoRA  & $83.8 \pm 0.6$ & $27.9 \pm 0.4$ \\
Simple two-view LoRA & $89.5 \pm 0.5$ & $24.4 \pm 0.3$ \\
\rowcolor{oursshade}
\textbf{\methodname{} paired 2N} & $\mathbf{90.5 \pm 0.4}$ & $\mathbf{28.6 \pm 0.3}$ \\
\methodname{}, high erased & $86.8 \pm 0.7$ & $24.3 \pm 0.4$ \\
\bottomrule
\end{tabular}
\end{table}

The same pattern appears on natural zero-shot checks, although the effect sizes
are smaller because these datasets are not constructed to isolate the upper
band. On InsectSet459, \methodname{} improves macro-AP over frozen AF-Next from 56.2 to 59.6 on all
clips and from 55.5 to 60.0 on high-frequency-rich clips. On LibriSeVoc, matched
rank improves from 66.6 to 68.7, while the erased control falls to 62.7. These
results suggest that the learned high-band path transfers beyond the synthetic
and counterfactual training format.

\subsection{Real Multiclass Zero-Shot Evaluation}

We also add datasets with multiclass answer formats, and Table~\ref{tab:multiclass}
summarizes the resulting transfer. On ESC-50 10-way classification, frozen
AF-Next reaches 75.5 BAcc. Low-band LoRA improves this to 83.8, showing that
part of the gain comes from answer-format adaptation. The full paired model
reaches $90.5 \pm 0.4$ BAcc across three seeds, and its high-erased control falls
to 86.8. Appendix~\ref{app:rain_case_study} shows one qualitative
rain/water example which confirms the importance of high-band evidence.

TinySOL is a useful negative result. The paired model (28.6 BAcc) is clearly above the
simple 2N baseline (24.4), but it does not clearly beat the low-band LoRA baseline (27.9).
This is consistent with isolated orchestral-note identity being mostly
available below 8~kHz.

\subsection{Downstream Fine-Tuning}

Generic wideband pretraining is also useful when the downstream task has both
low-band and high-band cues. On the Insect-10 test split (Section~\ref{sec:dataset}, Table~\ref{tab:insect10}), the low-band LoRA baseline
reaches 26.2 BAcc, indicating that the task has a low-band-solvable component.
\methodname{} reaches 55.4 BAcc after the same fine-tuning budget, while its
high-erased control falls to 22.7. The advantage over the baseline is therefore
again tied to the upper-band signal.

\begin{table}[t]
    \centering
    \caption{Insect-10 downstream fine-tuning results on the 113-item test split (10-way species classification over 10-second excerpts of InsectSet459 validation audio). Values are balanced accuracy in percent, averaged across seeds, after 300 fine-tuning steps for every model.}
    \label{tab:insect10}
    \small
    \begin{tabular}{lc}
        \toprule
        Model & Dev BAcc \\
        \midrule
        Low-band LoRA & $26.2$ \\
        \rowcolor{oursshade}
        \textbf{\methodname{} paired 2N} & $\mathbf{55.4}$ \\
        \methodname{} paired 2N, high-erased & $22.7$ \\
        \bottomrule
    \end{tabular}
\end{table}

We also tested downstream fine-tuning on the CEB bird-pair QA task
(Table~\ref{tab:ceb_bird}).
Low-band LoRA fine-tuning already improves over frozen AF-Next (64.36 to 72.28 mAP), so part of the task is solvable with frequencies below
8~kHz. Nevertheless, the paired wideband model obtains the best test balanced
accuracy, improving over simple 2N by 3.74 points, and drops by 5.05 points
under the high-erased control. The corresponding high-erased drop for simple 2N is
only 1.58 points. This suggests that the paired cross-frequency adapter uses
the 8--16~kHz stream more causally than a simple two-stream concatenation,
although the mAP gap between the paired and simple 2N models is small.

\begin{table}[t]
    \centering
    \caption{CEB bird-pair downstream fine-tuning test results. Values are percentages; $\pm$ reports SEM across seeds.}
    \label{tab:ceb_bird}
    \small
    \begin{tabular}{lcc}
        \toprule
        Model & Test BAcc ($\pm$ SEM) & Test mAP ($\pm$ SEM) \\
        \midrule
        Frozen AF-Next & $55.91 \pm 0.34$ & $64.36 \pm 0.72$ \\
        Low-band LoRA & $56.76 \pm 2.62$ & $72.28 \pm 0.78$ \\
        Simple two-view LoRA & $58.40 \pm 2.24$ & $74.93 \pm 0.88$ \\
        Simple two-view LoRA, high-erased & $56.82 \pm 3.19$ & $74.11 \pm 1.00$ \\
        \rowcolor{oursshade}
        \methodname{} paired 2N & $\mathbf{62.14 \pm 2.35}$ & $\mathbf{75.53 \pm 1.47}$ \\
        \methodname{} paired 2N, high-erased & $57.09 \pm 1.10$ & $73.36 \pm 1.81$ \\
        \bottomrule
    \end{tabular}
\end{table}

\subsection{Token Budget}

The main paired model uses $2N$ audio tokens, so we also test whether the same
quality can be retained with exactly $N$ tokens. Table~\ref{tab:token_budget}
summarizes the constructed-FSD development results. The half-rate exact-$N$ student,
trained with distillation from the $2N$ teacher on the paraphrase-rich
dataset, matches the teacher (83.6 vs.\ 83.3 mAP) while using half the audio tokens from the
first language-model layer onward.

\begin{table}[t]
    \centering
    \caption{Token-budget comparison on the constructed-FSD \emph{development} split. Note that this is a different subset from the sealed test split in Table~\ref{tab:constructed_fsd_test}, so absolute values are not directly comparable. Values are percentages; $\pm$ reports SEM across seeds where available.}
    \label{tab:token_budget}
    \small
    \begin{tabular}{llccc}
        \toprule
        Tokens & Method & BAcc & mAP & Rank \\
        \midrule
        $N$ & Frozen AF-Next & $48.3 \pm 1.1$ & $52.2 \pm 0.9$ & $51.7 \pm 1.8$ \\
        $2N$ & Simple two-view LoRA & $62.2 \pm 2.3$ & $76.0 \pm 0.8$ & $76.1 \pm 1.2$ \\
        \rowcolor{oursshade}
        $2N$ & \methodname{} paired & $71.9 \pm 1.5$ & $83.3 \pm 0.6$ & $80.6 \pm 1.4$ \\
        \rowcolor{oursshade}
        $N$ & Half-rate student & $\mathbf{72.2 \pm 1.1}$ & $\mathbf{83.6 \pm 0.8}$ & $\mathbf{82.2 \pm 1.5}$ \\
        \bottomrule
    \end{tabular}
\end{table}

We also tested whether the dedicated high-token block can be removed entirely,
keeping only adapter-enriched low tokens. This variant does not recover the
paired model's performance under either direct training or the distillation
recipe used for the half-rate model.

\section{Discussion}
\label{sec:discussion}

The experiments support three main conclusions (the ablations are summarized in Appendix~\ref{sec:ablations}, Table~\ref{tab:ablations}). First, a frozen 16~kHz encoder
can be reused for higher frequencies if the upper band is translated into its
input range and presented as a separate time-aligned stream. The high-erased
evaluations show that the gains are not merely caused by
extra tokens, prompt changes, or LoRA adaptation.

Second, we showed that paired local interaction is beneficial. The simple two-view LoRA baseline receives
the same low and high streams, but it lacks the timestamp-local adapter that
compares the bands before projection. Its weaker performance across the benchmarks
suggests that the model benefits from explicitly aligning low-band context with
upper-band evidence before the language model sees the audio prefix.

Third, the wideband gain does not require doubling the audio-token budget. The
half-rate exact-$N$ student matches the $2N$ model on constructed-FSD
development, although dedicated high-band token slots remain necessary.

\section{Limitations}
\label{sec:limitations}

The method currently targets the 8--16~kHz band. Audio sampled at 44.1 or
48~kHz can contain additional information above 16~kHz, which is not part of
the main \methodname{} path. 

We also test the method only using the AF-Next model. The approach is general, but in this work we do not explore the transferability of \methodname{} to other frozen audio-language models.

Adding a high-band path could in principle hurt ordinary low-band audio understanding. On the datasets reported in this paper, we verified that \methodname{} does not fall below the frozen base model (e.g., ESC-50 and TinySOL in Table~\ref{tab:multiclass}), but a systematic study of quality retention on broader speech, music, and general-audio benchmarks is left for future work.

Finally, the training dataset is synthetic or transformation-based by
design. This gives exact labels and helps avoid semantic overlap with
evaluation, but it also creates shortcut risks. Enriching the dataset with more real-world examples can lead to better results.

\section{Conclusion}
\label{sec:conclusion}

We introduced \methodname{}, a lightweight way to add 8--16~kHz information to
a frozen AF-Next model. The method keeps the original AF-Whisper encoder and
language model frozen, translates the upper band into the encoder's passband,
and uses timestamp-local paired attention to fuse low- and high-band information.
Across counterfactual, natural, and downstream evaluations, the model learns to
use high-frequency information that is invisible to the stock 16~kHz pipeline.
The strongest exact-$N$ variant further shows that wideband transfer does not
necessarily require doubling the audio-token budget.

\bibliographystyle{nebius}
\bibliography{references}

\begin{thebibliography}{23}
\providecommand{\natexlab}[1]{#1}
\providecommand{\url}[1]{\texttt{#1}}
\expandafter\ifx\csname urlstyle\endcsname\relax
  \providecommand{\doi}[1]{doi: #1}\else
  \providecommand{\doi}{doi: \begingroup \urlstyle{rm}\Url}\fi

\bibitem[{CEB Train-XenoCanto Contributors}(2026)]{ceb2026xenocanto}
{CEB Train-XenoCanto Contributors}.
\newblock {CEB} train-xenocanto high-frequency bird events.
\newblock \url{https://zenodo.org/records/20762099}, 2026.
\newblock Accessed 2026-08-27.

\bibitem[Cella et~al.(2020)Cella, Ghisi, Lostanlen, Levy, Fineberg, and Maresz]{cella2020tinysol}
Cella, C.-E., Ghisi, D., Lostanlen, V., Levy, F., Fineberg, J., and Maresz, Y.
\newblock {TinySOL}: An audio dataset of isolated musical notes, 2020.

\bibitem[Chu et~al.(2023)Chu, Xu, Zhou, Yang, Zhang, Yan, Zhou, and Zhou]{chu2023qwenaudio}
Chu, Y., Xu, J., Zhou, X., Yang, Q., Zhang, S., Yan, Z., Zhou, C., and Zhou, J.
\newblock {Qwen-Audio}: Advancing universal audio understanding via unified large-scale audio-language models, 2023.

\bibitem[Chu et~al.(2024)Chu, Xu, Yang, Wei, Wei, Guo, Leng, Lv, He, Lin, Zhou, and Zhou]{chu2024qwen2audio}
Chu, Y., Xu, J., Yang, Q., Wei, H., Wei, X., Guo, Z., Leng, Y., Lv, Y., He, J., Lin, J., Zhou, C., and Zhou, J.
\newblock {Qwen2-Audio} technical report, 2024.

\bibitem[Fai{\ss} et~al.(2025)Fai{\ss}, Ghani, and Stowell]{faiss2025insectset459}
Fai{\ss}, M., Ghani, B., and Stowell, D.
\newblock {InsectSet459}: An open dataset of insect sounds for bioacoustic machine learning, 2025.

\bibitem[Fang et~al.(2026)Fang, Gao, Li, Liu, Zhong, and Liu]{fang2026amsa}
Fang, X., Gao, T., Li, R., Liu, H., Zhong, Y., and Liu, J.
\newblock {AMSA}: Anchored multi-sampling-rate adaptation via subband tokenization for audio foundation models, 2026.
\newblock Preprint.

\bibitem[Fonseca et~al.(2022)Fonseca, Favory, Pons, Font, and Serra]{fonseca2022fsd50k}
Fonseca, E., Favory, X., Pons, J., Font, F., and Serra, X.
\newblock {FSD50K}: An open dataset of human-labeled sound events.
\newblock \emph{IEEE/ACM Transactions on Audio, Speech, and Language Processing}, 30:\penalty0 829--852, 2022.
\newblock \doi{10.1109/TASLP.2021.3133208}.

\bibitem[Gemmeke et~al.(2017)Gemmeke, Ellis, Freedman, Jansen, Lawrence, Moore, Plakal, and Ritter]{gemmeke2017audioset}
Gemmeke, J.~F., Ellis, D. P.~W., Freedman, D., Jansen, A., Lawrence, W., Moore, R.~C., Plakal, M., and Ritter, M.
\newblock {Audio Set}: An ontology and human-labeled dataset for audio events.
\newblock In \emph{Proceedings of the IEEE International Conference on Acoustics, Speech and Signal Processing}, pp.\  776--780, 2017.
\newblock \doi{10.1109/ICASSP.2017.7952261}.

\bibitem[Ghosh et~al.(2025)Ghosh, Kong, Kumar, Sakshi, Kim, Ping, Valle, Manocha, and Catanzaro]{ghosh2025audioflamingo2}
Ghosh, S., Kong, Z., Kumar, S., Sakshi, S., Kim, J., Ping, W., Valle, R., Manocha, D., and Catanzaro, B.
\newblock {Audio Flamingo 2}: An audio-language model with long-audio understanding and expert reasoning abilities, 2025.

\bibitem[Ghosh et~al.(2026)Ghosh, Goel, Jayakumar, Koroshinadze, Anand, Kong, Gururani, Lee, Kim, Aljafari, Yang, Kim, Duraiswami, Manocha, Shoeybi, Catanzaro, Liu, and Ping]{ghosh2026afnext}
Ghosh, S., Goel, A., Jayakumar, K., Koroshinadze, L., Anand, N., Kong, Z., Gururani, S., Lee, S.-g., Kim, J., Aljafari, A., Yang, C.-H.~H., Kim, S., Duraiswami, R., Manocha, D., Shoeybi, M., Catanzaro, B., Liu, M.-Y., and Ping, W.
\newblock {Audio Flamingo Next}: Next-generation open audio-language models for speech, sound, and music, 2026.

\bibitem[Goel et~al.(2025)Goel, Ghosh, Kim, Kumar, Kong, Lee, Yang, Duraiswami, Manocha, Valle, and Catanzaro]{goel2025audioflamingo3}
Goel, A., Ghosh, S., Kim, J., Kumar, S., Kong, Z., Lee, S.-g., Yang, C.-H.~H., Duraiswami, R., Manocha, D., Valle, R., and Catanzaro, B.
\newblock {Audio Flamingo 3}: Advancing audio intelligence with fully open large audio language models, 2025.

\bibitem[Gong et~al.(2023)Gong, Luo, Liu, Karlinsky, and Glass]{gong2023ltu}
Gong, Y., Luo, H., Liu, A.~H., Karlinsky, L., and Glass, J.
\newblock {LTU}: A general-purpose large language model for speech, audio, and music, 2023.

\bibitem[Hu et~al.(2022)Hu, Shen, Wallis, Allen-Zhu, Li, Wang, Wang, and Chen]{hu2022lora}
Hu, E.~J., Shen, Y., Wallis, P., Allen-Zhu, Z., Li, Y., Wang, S., Wang, L., and Chen, W.
\newblock {LoRA}: Low-rank adaptation of large language models.
\newblock In \emph{International Conference on Learning Representations}, 2022.

\bibitem[Huang et~al.(2026)Huang, Ge, Wang, Li, Wang, Wang, and Dang]{huang2026msrhubert}
Huang, Z., Ge, M., Wang, T., Li, X., Wang, X., Wang, L., and Dang, J.
\newblock {MSR-HuBERT}: Self-supervised pre-training for adaptation to multiple sampling rates, 2026.

\bibitem[Kong et~al.(2024)Kong, Goel, Badlani, Ping, Valle, and Catanzaro]{kong2024audioflamingo}
Kong, Z., Goel, A., Badlani, R., Ping, W., Valle, R., and Catanzaro, B.
\newblock {Audio Flamingo}: A novel audio language model with few-shot learning and dialogue abilities, 2024.

\bibitem[Ly(2026)]{ly2026tinygiant}
Ly, V.-T.
\newblock {TinyGiantALM}: A compact audio-language model for intent-aware reasoning under resource constraints, 2026.

\bibitem[{NVIDIA}(2026)]{nvidia2026audioflamingonexthf}
{NVIDIA}.
\newblock {Audio Flamingo Next HF}.
\newblock \url{https://huggingface.co/nvidia/audio-flamingo-next-hf}, 2026.
\newblock Hugging Face model card; accessed 2026-09-02. License: NVIDIA OneWay Noncommercial License.

\bibitem[Piczak(2015)]{piczak2015esc}
Piczak, K.~J.
\newblock {ESC}: Dataset for environmental sound classification.
\newblock In \emph{Proceedings of the ACM International Conference on Multimedia}, pp.\  1015--1018, 2015.
\newblock \doi{10.1145/2733373.2806390}.

\bibitem[Radford et~al.(2022)Radford, Kim, Xu, Brockman, McLeavey, and Sutskever]{radford2022whisper}
Radford, A., Kim, J.~W., Xu, T., Brockman, G., McLeavey, C., and Sutskever, I.
\newblock Robust speech recognition via large-scale weak supervision, 2022.

\bibitem[Shannon(1949)]{shannon1949communication}
Shannon, C.~E.
\newblock Communication in the presence of noise.
\newblock \emph{Proceedings of the IRE}, 37\penalty0 (1):\penalty0 10--21, 1949.
\newblock \doi{10.1109/JRPROC.1949.232969}.

\bibitem[Sun et~al.(2023)Sun, Jia, Hou, and Lyu]{sun2023librisevoc}
Sun, C., Jia, S., Hou, S., and Lyu, S.
\newblock Ai-synthesized voice detection using neural vocoder artifacts, 2023.

\bibitem[Svendsen et~al.(2026)]{svendsen2026seal}
Svendsen, L.~M. et~al.
\newblock Hydraulic seal wear classification by fine-tuning a transformer-based audio model using acoustic emission.
\newblock \emph{Sensors}, 26\penalty0 (9):\penalty0 2856, 2026.
\newblock \doi{10.3390/s26092856}.

\bibitem[Tang et~al.(2023)Tang, Yu, Sun, Chen, Tan, Li, Lu, Ma, and Zhang]{tang2023salmonn}
Tang, C., Yu, W., Sun, G., Chen, X., Tan, T., Li, W., Lu, L., Ma, Z., and Zhang, C.
\newblock {SALMONN}: Towards generic hearing abilities for large language models, 2023.

\end{thebibliography}
\newpage
\appendix
\begin{center}
    {\nebiusheadfont\Large\color{nebiusdeepblue}Appendix}
\end{center}

\section{Dataset Examples}
\label{app:dataset_examples}

This appendix gives representative examples of the question formats used in
training, downstream fine-tuning, and evaluation. These are examples of the
templates and task semantics rather than additional evaluation results.

\subsection{Generic High-Frequency Training}

The generic training dataset avoids downstream sound classes and instead asks about
signal properties whose labels are known from the applied transformation. For a
bandwidth-quality item, the audio is a real recording that may have been
low-pass filtered:
\begin{nebiusexample}
Does this recording retain its natural crisp top end, or were only its highest
frequencies removed?\\
A) natural full bandwidth \quad B) highest frequencies removed
\end{nebiusexample}

\begin{nebiusexample}
The same scene is played twice with a pause. Which copy is free of an added
steady hiss in its top octave?\\
A) the first copy \quad B) the second copy
\end{nebiusexample}

Low-band-only examples deliberately include high-band distractors so the
model does not learn that the upper band is always relevant:
\begin{nebiusexample}
Compare the two playbacks: which carries more energy under 8~kHz?\\
A) the first copy \quad B) the second copy
\end{nebiusexample}
In these examples, the two copies can have different borrowed upper bands, but
the label depends only on the 0--8~kHz content.

We also add multiclass training examples. In quarter localization, the
top-octave energy is strongest in exactly one quarter of the clip:
\begin{nebiusexample}
Split the clip into four equal parts. Which part carries the strongest energy
above 8~kHz?\\
A) first quarter \quad B) second quarter \quad C) third quarter \quad
D) fourth quarter
\end{nebiusexample}

\subsection{Downstream Fine-Tuning}

Insect-10 is the 10-way species-classification task constructed from
InsectSet459 validation audio as described in Section~\ref{sec:dataset}. A representative prompt is:
\begin{nebiusexample}
Which insect species is producing the main sound in this recording?\\
A) \emph{Cicada orni} \quad \ldots \quad
J) \emph{Chorthippus biguttulus}
\end{nebiusexample}
The split is observation-disjoint, and models fine-tuned on this suite are not
used for the corresponding insect zero-shot claims.

The CEB bird-pair downstream suite uses binary species-pair questions rather
than the full 87-way taxonomy because the larger classification task was partly
solvable from low-band cues. A representative item asks:
\begin{nebiusexample}
Which bird species is producing the main high-frequency vocalization in this
recording?\\
A) species from pair member one \quad B) species from pair member two
\end{nebiusexample}
The source-recording-separated split contains balanced labels and answer
letters, and is used only for downstream fine-tuning evidence.

\subsection{Zero-Shot Evaluation}

The constructed-FSD benchmark asks semantic questions whose answer differs only
through the 8--16~kHz donor signal:
\begin{nebiusexample}
Does this recording contain a cymbal or hi-hat sound?\\
A) no cymbal or hi-hat sound \quad
B) yes, a cymbal or hi-hat sound is present
\end{nebiusexample}
The positive and negative pair members share the same low-band carrier and
differ only in the energy-matched upper-band donor.

Natural zero-shot checks use published datasets without training-task overlap.
For InsectSet459, the model chooses a broad insect group:
\begin{nebiusexample}
Which broad type of insect is producing the main sound in this recording?\\
A) a grasshopper, cricket, or katydid \quad B) a cicada
\end{nebiusexample}
For LibriSeVoc, the task asks whether genuine speech ranks above vocoder
resynthesis:
\begin{nebiusexample}
Is this speech an original human recording or a waveform generated by a neural
vocoder?\\
A) neural-vocoder generated recording \quad B) genuine human recording
\end{nebiusexample}

The real multiclass benchmarks use ordinary audio-understanding
formats. ESC-50 asks which environmental sound is present among ten
same-supercategory options. TinySOL asks which instrument plays an isolated
note. These examples test whether the wideband path adds high-frequency
sensitivity on standard classification tasks.

\section{Rain/Water Example}
\label{app:rain_case_study}

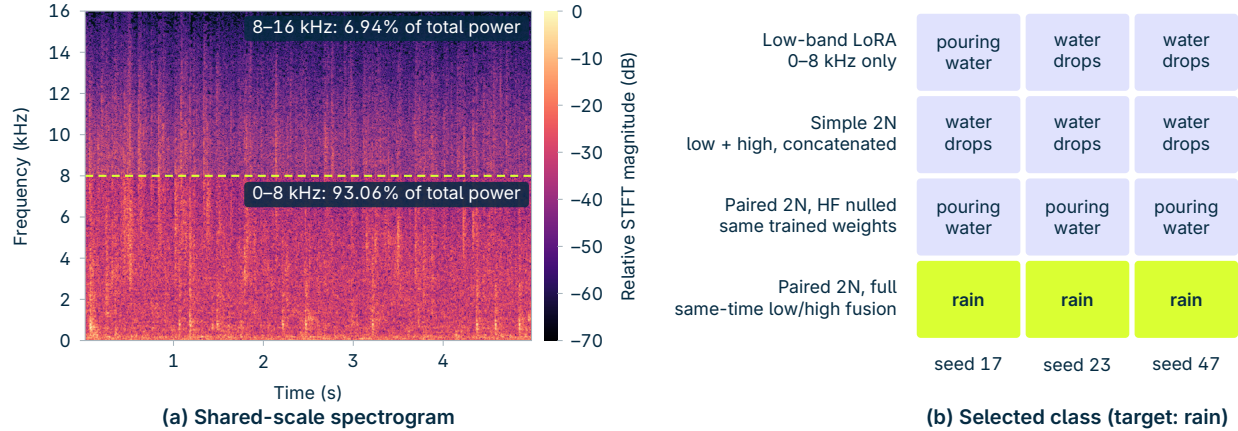
\begin{figure}[t]
    \centering
    \begingroup
\nebiussans
\pgfplotsset{colormap={magma}{%
  rgb255=(0,0,3) rgb255=(3,3,15) rgb255=(9,7,32) rgb255=(18,13,49)
  rgb255=(28,16,68) rgb255=(39,18,88) rgb255=(52,16,105) rgb255=(66,15,117)
  rgb255=(79,18,123) rgb255=(92,22,127) rgb255=(104,28,129) rgb255=(117,33,129)
  rgb255=(129,37,129) rgb255=(142,42,129) rgb255=(156,46,127) rgb255=(170,51,125)
  rgb255=(183,55,121) rgb255=(194,59,117) rgb255=(207,64,112) rgb255=(219,71,106)
  rgb255=(229,80,100) rgb255=(238,91,94) rgb255=(244,105,92) rgb255=(249,120,93)
  rgb255=(251,135,97) rgb255=(253,150,104) rgb255=(254,165,113) rgb255=(254,180,123)
  rgb255=(254,196,136) rgb255=(254,211,149) rgb255=(253,226,163) rgb255=(252,240,178)
  rgb255=(252,253,191)}}
\begin{tikzpicture}[
  font=\scriptsize,
  color=nebiusdeepblue,
  cell/.style={rounded corners=1.5pt, minimum width=39pt, minimum height=29pt,
               inner sep=0pt, align=center, font=\scriptsize, text=nebiusdeepblue},
  wrong/.style={cell, fill=wrongfill},
  correct/.style={cell, fill=correctfill, font=\nebiussemibold\scriptsize},
  powerlabel/.style={anchor=east, fill=nebiusdeepblue, fill opacity=0.85,
                     text opacity=1, text=white, rounded corners=1.5pt,
                     inner xsep=3pt, inner ysep=1.6pt, font=\scriptsize},
]
\begin{axis}[
  nebius image axis,
  name=spec,
  at={(0,0)}, anchor=south west,
  scale only axis, width=168pt, height=124pt,
  xmin=0.0175, xmax=4.9825, ymin=0, ymax=16,
  xtick={1,2,3,4}, ytick={0,2,...,16},
  xlabel={Time (s)}, ylabel={Frequency (kHz)},
  colormap name=magma,
  point meta min=-70, point meta max=0,
  colorbar right,
  colorbar style={
    width=5pt,
    xshift=-4pt,
    axis line style={draw=none},
    ytick={-70,-60,...,0},
    yticklabels={\textendash 70,\textendash 60,\textendash 50,\textendash 40,
                 \textendash 30,\textendash 20,\textendash 10,0},
    ytick pos=right,
    tick align=outside,
    ytick style={draw=nebiusdeepblue!35,line width=0.4pt},
    major tick length=2pt,
    tick label style={font=\nebiusfont\scriptsize,text=nebiusdeepblue},
    ylabel={Relative STFT magnitude (dB)},
    ylabel style={font=\nebiusfont\scriptsize,text=nebiusdeepblue},
  },
]
\addplot graphics[xmin=0.0175, xmax=4.9825, ymin=0, ymax=16]
  {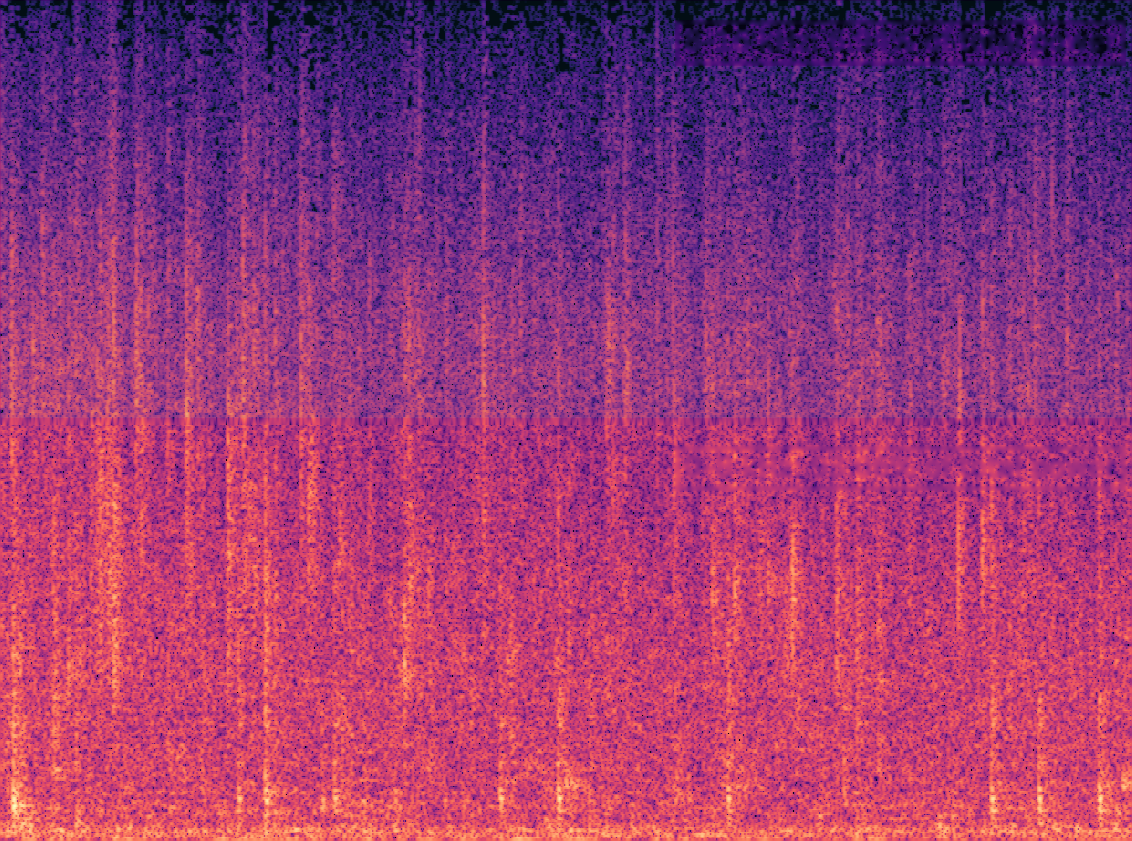};
\draw[nebiuslime, line width=0.8pt, dash pattern=on 3pt off 2pt]
  (axis cs:0.0175,8) -- (axis cs:4.9825,8);
\node[powerlabel] at (axis cs:4.96,15.16) {8\textendash 16 kHz: 6.94\% of total power};
\node[powerlabel] at (axis cs:4.96,7.09) {0\textendash 8 kHz: 93.06\% of total power};
\end{axis}

\begin{scope}[shift={(312pt,0pt)}]
  \foreach \y/\model in {%
      108.5/{Low-band LoRA\\ 0\textendash 8 kHz only},
      77.5/{Simple 2N\\ low + high, concatenated},
      46.5/{Paired 2N, HF nulled\\ same trained weights},
      15.5/{Paired 2N, full\\ same-time low/high fusion}}{%
    \node[anchor=east, align=right] at (-3pt, \y pt) {\model};
  }
  \foreach \x/\seed in {20.5/17, 61.5/23, 102.5/47}{%
    \node[anchor=north] at (\x pt, -3pt) {seed \seed};
  }
  \node[wrong] at (20.5pt, 108.5pt)  {pouring\\ water};
  \node[wrong] at (61.5pt, 108.5pt)  {water\\ drops};
  \node[wrong] at (102.5pt, 108.5pt) {water\\ drops};
  \node[wrong] at (20.5pt, 77.5pt)  {water\\ drops};
  \node[wrong] at (61.5pt, 77.5pt)  {water\\ drops};
  \node[wrong] at (102.5pt, 77.5pt) {water\\ drops};
  \node[wrong] at (20.5pt, 46.5pt)  {pouring\\ water};
  \node[wrong] at (61.5pt, 46.5pt)  {pouring\\ water};
  \node[wrong] at (102.5pt, 46.5pt) {pouring\\ water};
  \node[correct] at (20.5pt, 15.5pt)  {rain};
  \node[correct] at (61.5pt, 15.5pt)  {rain};
  \node[correct] at (102.5pt, 15.5pt) {rain};
\end{scope}

\node[anchor=north, font=\nebiussemibold\footnotesize]
  at ($(spec.south)+(0,-22pt)$) {(a) Shared-scale spectrogram};
\node[anchor=north, font=\nebiussemibold\footnotesize]
  at ($(312pt+61.5pt,0)+(0,-22pt)$) {(b) Selected class (target: rain)};
\end{tikzpicture}
\endgroup
    \caption{Qualitative ESC-50 case study. Only the paired same-time low/high fusion model
    consistently selects the correct target class across three seeds. Nulling the
    high-frequency view changes the paired model's prediction while keeping the
    checkpoint and token layout fixed.}
    \label{fig:rain_water_case}
\end{figure}

Figure~\ref{fig:rain_water_case} shows a heavy-rain recording from ESC-50
(\texttt{5-203739-A-10}), where 6.94\% of the signal power lies between 8 and
16~kHz. Across three training seeds, the low-band LoRA model confuses the clip
with pouring water or water drops. The simple 2N model also receives the
translated upper band, but only as a concatenated second stream, and predicts
water drops for all three seeds. In contrast, the paired 2N model, which fuses
low- and high-band tokens at matching timestamps, identifies rain for all three
seeds. When only the paired model's 8--16~kHz view is nulled while preserving
the same trained weights and 2N token layout, all three predictions change to
pouring water. This example shows the importance of high-band information
as well as of the timestamp-local fusion mechanism, especially when two classes
are very similar to each other but still distinguishable by a human listener.

\section{Ablations}
\label{sec:ablations}

We run ablations to separate four possible sources of improvement: task-format
adaptation, access to the high-frequency stream, the paired cross-band adapter,
and the extra token budget. These controls are important because a higher score
from the full model alone would not show which part of the system is actually
responsible for the gain.

\begin{table}[t]
    \centering
    \caption{Summary of the main ablations and what each one tests.}
    \label{tab:ablations}
    \small
    \begin{tabular}{p{0.25\linewidth}p{0.35\linewidth}p{0.30\linewidth}}
        \toprule
        Ablation & What we change & What it shows \\
        \midrule
        Low-band LoRA & Train the same LoRA modules, but provide only the stock low-band audio tokens. & Separates answer-format and prompt adaptation from true high-frequency use. \\
        Simple 2N & Concatenate low and translated-high streams without paired cross-band attention. & Tests whether merely exposing the upper band is enough. \\
        High-erased control & Keep the same trained checkpoint and token layout, but replace the high stream by digital silence. & Measures the causal contribution of the 8--16~kHz waveform. \\
        Paired residual disabled & Keep both streams but turn off the learned paired residual. & Isolates the value of timestamp-local cross-frequency interaction. \\
        Exact-$N$ compression & Reduce the wideband prefix from $2N$ to $N$ tokens by half-rate pooling. & Tests whether the high-band gain requires doubling the audio-token budget. \\
        No high-token block & Keep only the adapter-enriched low tokens and remove dedicated high-band slots. & Tests whether high evidence can be fully folded into the low stream. \\
        \bottomrule
    \end{tabular}
\end{table}

The low-band LoRA and simple-2N controls answer different questions. Low-band
LoRA shows how much improvement comes from learning the multiple-choice format
and the training distribution without receiving any high-frequency signal.
Simple 2N receives the translated high stream, but does not force low and high
states at the same timestamp to interact before projection. On constructed-FSD,
simple 2N improves strongly over frozen AF-Next, but remains well below the
paired model. This suggests that access to the high band is necessary but not
sufficient: the model also benefits from an explicit local mechanism that
compares the two frequency views before they enter the language model.

The high-erased control is the main causal ablation. We replace the high stream with the null signal. On the
sealed constructed-FSD test, \methodname{} drops from 81.4 to 49.4 mAP under
this intervention. This drop shows that the gains are not explained by the
larger prefix, LoRA, or the changed input format alone.

We also ablate the paired residual itself. In this setting both the low and
high streams remain available, but the learned timestamp-local residual from
Eq.~\ref{eq:gate} is disabled. This removes only the cross-band interaction
while preserving the rest of the two-stream layout. Disabling the residual
degrades development-set performance, showing that \methodname{} is not simply a concatenation
method; the learned local comparison between the low-band context and
high-minus-null evidence contributes additional information.

The token-budget ablations test whether high-frequency transfer requires a
larger audio prefix. With distillation on
the paraphrase-rich dataset, the half-rate student matches the $2N$ teacher
on constructed-FSD development. However, removing the high-token block entirely
does not recover the paired model's performance under either direct training or the distillation recipe. Thus the high
band still needs dedicated token capacity, but $N/2$ high-band slots are enough
in the current exact-$N$ design. We note that the pooling module was distilled
only on the binary (two-option) questions of the training dataset, so the
half-rate student has been validated mainly on binary and paired tasks; its
behavior on multiclass formats is less constrained and remains to be studied.

\end{document}